\documentclass[fleqn,usenatbib]{mnras} 

\usepackage{natbib}
\usepackage{newtxtext,newtxmath}
\usepackage{stfloats}
\usepackage[flushleft]{threeparttable}

\usepackage[T1]{fontenc}
\usepackage{setspace}
\DeclareRobustCommand{\VAN}[3]{#2}
\let\VANthebibliography\thebibliography
\def\thebibliography{\DeclareRobustCommand{\VAN}[3]{##3}\VANthebibliography}

\usepackage{graphicx}	
\usepackage{epstopdf}
\usepackage{amsmath}	

\usepackage{amssymb}	
\usepackage[justification=centering]{caption}
\usepackage{float} 
\usepackage{courier} 

\usepackage{color,xcolor}

\title[QPOs in the CDF-S]{Searching for Quasi-periodic Oscillations in Active Galactic Nuclei of the Chandra Deep Field South}
\author[Bao \& Li]{
Tong Bao$^{1,2}$\thanks{E-mail: baotong@smail.nju.edu.cn},
Zhiyuan Li$^{1,2}$\thanks{E-mail: lizy@nju.edu.cn}
\\
$^{1}$School of Astronomy and Space Science, Nanjing University, Nanjing 210046, China\\
$^{2}$Key Laboratory of Modern Astronomy and Astrophysics (Nanjing University), Ministry of Education, Nanjing 210046, China
}

\date{Accepted XXX. Received YYY; in original form ZZZ}

\pubyear{2015}

\begin{document}
\label{firstpage}
\pagerange{\pageref{firstpage}--\pageref{lastpage}}
\maketitle

\begin{abstract}
Recent X-ray observations have revealed 
growing evidence of quasi-periodic oscillation (QPO) in the light curve of active galactic nuclei (AGNs), which may serve as a useful probe of black hole physics.
In this work, we present a systematic search for X-ray QPOs among $\sim$1000 AGNs of the {\it Chandra} Deep Field South (CDF-S) in a homogeneous fashion.
Dividing the 7-Ms {\it Chandra} observations into four epochs, we search for periodic signals that are persistent throughout any of these epochs, using two independent methods: Lomb-Scargle periodogram and Gregory-Loredo Algorithm.
No statistically significant periodic signal is found with either method on any of the four epochs.
Our extensive simulations of source light curves suggest that this non-detection is primarily due to a moderate sensitivity of the CDF-S data in QPO detection.  
Using the simulation-predicted detection efficiency, we are able to provide a meaningful constraint on the intrinsic occurrence rate of persistent QPOs,  $< (15-20)\%$, provided that they share a similar power spectral density with a handful of currently known AGN QPOs. 
The true intrinsic occurrence rate might be significantly below this upper limit, however, given the non-detection among the CDF-S sources.
Our additional search for short-lived QPOs that are only detected over a small subset of all observations results in two candidates, one in source XID 643 at a period of $\sim$13273 s and the other in source XID 876 at a period of $\sim$7065 s. 
\end{abstract}

\begin{keywords}
(galaxies:) quasars: supermassive black holes -- galaxies: nuclei -- X-rays: galaxies
\end{keywords}



\section{Introduction}


The phenomenon of X-ray quasi-periodic oscillation (QPO) has long been known for many kinds of accreting systems, especially in black hole binaries (BHBs).  Generally, BHBs exhibit two classes of QPOs \citep{2019NewAR..8501524I}, namely, the low-frequency QPOs (LFQPOs, $\nu \sim $ 0.1--10 Hz) and high-frequency QPOs (HFQPOs, $\nu \sim$ hundreds of Hz).
The exact origin of QPOs is still an open question; leading scenarios include Lens–Thirring precession for the LFQPOs \citep{1999ApJ...524L..63S} and Keplerian motion near the innermost stable circular orbit (ISCO) of the BH for the HFQPOs \citep{2006ARA&A..44...49R}.
The latter case, in particular, suggests that QPOs may provide an important probe of the immediate vicinity of black holes.  

X-ray QPOs have also been found in a growing number of active galactic nuclei (AGNs). 
The first robust detection of AGN QPO was a 1-hour modulation from the narrow-line type-1 Seyfert (NLS1) galaxy RE J1034+396 \citep{2008Natur.455..369G}, which has a quality factor $Q \equiv \nu/ \delta \nu  \approx 15$ and a fractional root-mean-square (rms) variability $R_{\rm rms} \approx 4.7\%$.  
After that, several more X-ray QPOs in NLS1s have been reported, which include MS 2254.9-3712 ($P = 1/\nu \sim$2 hour, $Q \sim 8$ and $R_{\rm rms} \sim 6\%$, \citealp{2015MNRAS.449..467A}), Mrk 766 ($P \sim$ 1.8 hour, $Q > 13.6$ and $R_{\rm rms} \sim 14.3\%$, \citealp{2017ApJ...849....9Z}), 1H 0707-495 ($P \sim$ 1 hour, $Q > 15$ and $R_{\rm rms} \sim 15\%$, \citealp{2016ApJ...819L..19P}; $P \sim$ 2.3 hour, \citealp{2018ApJ...853..193Z}), and MCG-06-30-15 ($P \sim$ 1 hour, \citealp{2018A&A...616L...6G}). 
QPOs have also been found in type-2 Seyfert galaxies, which include 2XMM J123103.2+110648 ($P \sim$ 3.8 hour, $Q > 5$ and $R_{\rm rms} \sim 25-50\%$, \citealp{2013ApJ...776L..10L}) and XMMU J134736+173403 ($P \sim$ 23.8 hour, \citealp{2018MNRAS.477.3178C}).
A notable aspect of AGN QPOs, at least in some sources, is their intermittency. 
For instance, the 1-hour QPO in 1H 0707–495 was detected in only one XMM-Newton observation lasting 55 ks, while absent in the other 14 observations  of 1145 ks total exposure \citep{2016ApJ...819L..19P}; 
the 3.8-hour QPO in 2XMM J123103.2+110648 was detected with two observations in 2005 but not with the one in 2003 \citep{2013ApJ...776L..10L};
RE J1034+396 exhibited the 1-hour QPO in six out of eight XMM-Newton observations between 2002--2018 \citep{2020MNRAS.495.3538J}. 

More recently, a peculiar class of quasi-periodic flux variation, called quasi-periodic eruption (QPE), emerges.
The first such case was found in the X-ray light curve of GSN 069, with high-amplitude (up to $\sim$ 100) flares separated by a period of $\sim 9$ hours \citep{2019Natur.573..381M}.
A similar case is seen in RX J1301.9+274715, which exhibited high-amplitude soft-X-ray flares recurring quasi-periodically on a timescale of 13–20 ks \citep{2020A&A...636L...2G}. 
Moreover, \citet{2020A&A...644L...9S} found that this source exhibited a 0.4 hour QPO in the quiescent state, as caught by two XMM-Newton observations taken respectively in 2000 and 2019.

The recent advent of the {\it eROSITA} all-sky survey led to the discovery of new QPEs in two otherwise quiescent galaxies, named eRO-QPE1 and eRO-QPE2, with a period of 18.5 hours and 2.4 hours, respectively \citep{2021Natur.592..704A}. 
The origin of QPEs and their relation with normal QPOs are not clear.
Notably, in the literature there are also claims of short-period QPOs associated with tidal disruption events, such as that in Swift J164449.3+573451 ($P \sim$ 200 s, \citealp{2012Sci...337..949R}) and ASASSN-14li ($P \sim$ 131 s, \citealp{2019Sci...363..531P}), which appear to be distinct from the normal AGN QPOs \citep{2021ApJ...906...92S}.

Given their frequency range of $10^{-4}-10^{-2}$ Hz, the AGN QPOs are naturally understood as the counterpart of BHB HFQPOs, 
based on the {\it naive} assumption that the QPO period is primarily scaled with black hole mass ($M_{\rm BH} \sim 10^6-10^8{\rm~M_\odot}$ for super-massive black holes versus $M_{\rm BH} \sim 10{\rm~M_\odot}$ for stellar-mass black holes), 
although the black hole spin must also be relevant provided that these QPOs are originated near the ISCO.
The same argument would put the AGN counterpart of BHB LFQPO at an expected quasi-period of days to months.
QPOs with such a long period (low frequency), however, are difficult to detect with the existed AGN monitoring campaigns \citep{2005MNRAS.362..235V}. 

Thus far, the majority of AGN QPOs were discovered serendipitously. Perhaps the only dedicated search for AGN QPOs, to our knowledge, is the work of \citet{2012A&A...544A..80G}, who identified only one QPO signal (the previously known REJ1034+396) among 104 bright nearby ($z < 0.4$) AGNs with available XMM-Newton observations.
This apparently low detection rate of AGN QPOs is similar to the case of BHB HFQPOs \citep{2012MNRAS.426.1701B}, posing an interesting question about the intrinsic occurrence rate and duty cycle of AGN QPOs. 
We are thus motivated to perform the first systematic search for AGN QPOs in deep X-ray surveys, focusing in particular the 7-Ms {\it Chandra} Deep Field-South (CDF-S; \citealp{2017ApJS..228....2L}). 
Deep surveys like the CDF-S have the advantage of a nearly unbiased coverage of hundreds of AGNs in the same time, allowing for an efficient and homogeneous QPO search. 
Moreover, the deep fields typically consist of repeated observations spanning a long temporal baseline, which are well suited to examine the likelihood of AGN QPOs being persistent (or intermittent).
Last but nor least, multi-wavelength information are often available for the deep field sources, which can help determine the AGN host properties, when a QPO is detected. 

This work is organized as follows. In Section \ref{sec:obs}, we describe our merit and procedure in preparing source light curves from the {\it Chandra} observations. 
In Section~\ref{sec:timing}, we introduce the QPO searching methods and present the results. 
Simulations of AGN light curve, designed to assess the sensitivity and detection efficiency of the period-searching method, are described in Section \ref{sec:simulation}.
In Section \ref{sec:discussion}, we employ the simulation prediction to constrain the intrinsic occurrence rate of persistent AGN QPOs. We also examine the possible existence of transient QPOs. A brief summary of our study is given in Section \ref{sec:summary}.

\section{{ \it Chandra} observations and Data preparation}
\label{sec:obs} 

The 7-Ms CDF-S, the deepest X-ray survey of distant AGNs and active galaxies ever conducted, is composed of 102 individual {\it Chandra} observations, all taken by the Advanced CCD Imaging Spectrometer (ACIS) with its I-array providing the primary field-of-view (FoV) (see Figure 1 in \citealp{2017ApJS..228....2L}). 

\begin{figure}
 
 \centering
\includegraphics[scale=0.41]{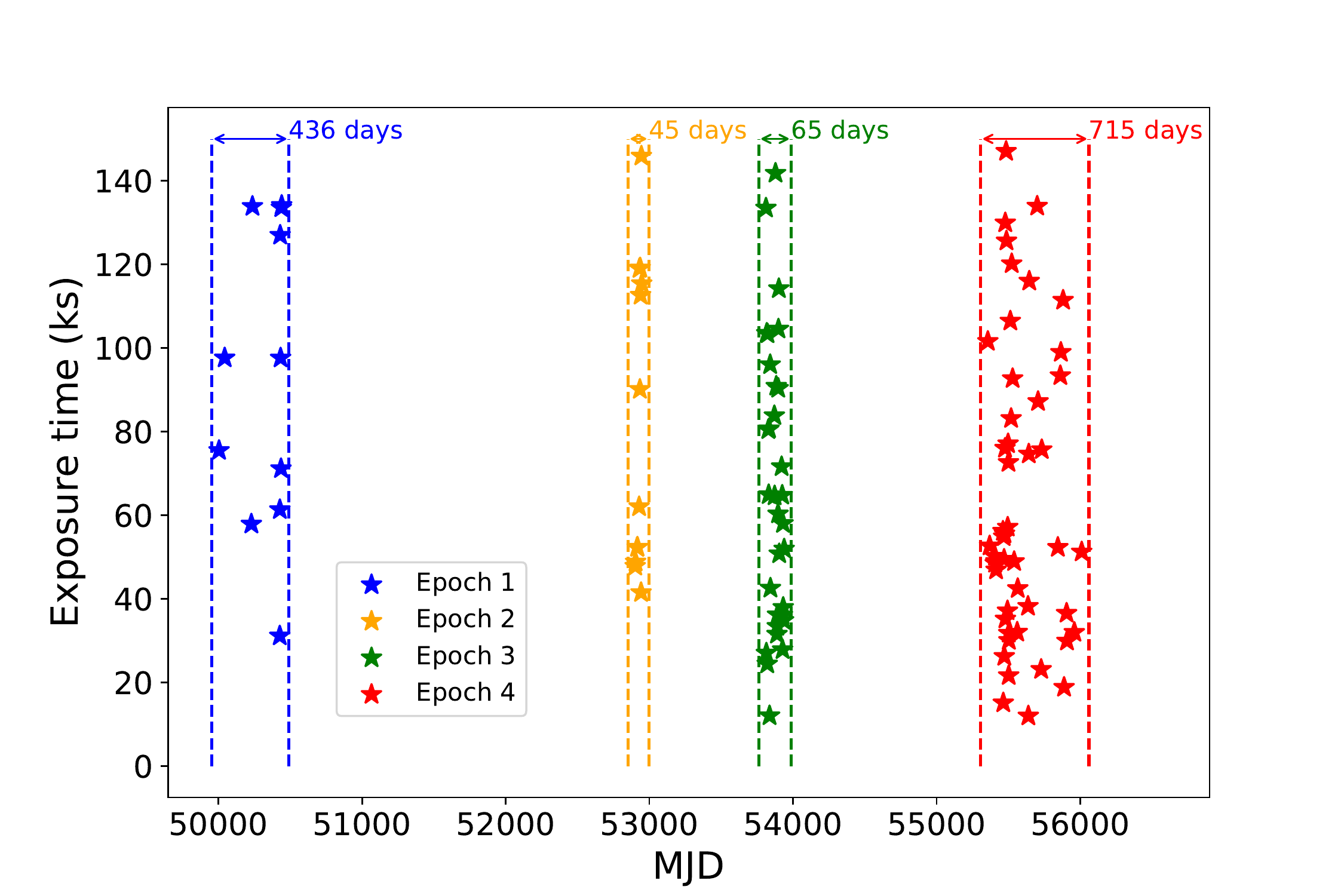}

\caption{An illustration of the observing cadence of the CDF-S. Four epochs of closely separated observations are color-coded, within which QPOs are searched for. The start and end dates of each epoch are marked by a pair of dashed lines.}

\label{fig:Epochobs}
 	
 \end{figure}

The 102 ACIS-I observations together cover a temporal baseline of 16.5 years, but with a highly irregular cadence. 
Wide gaps between successive observations can lead to formidably large computational effort in timing analysis.
Therefore, for our purpose of QPO searching, we group these observations into four effective epochs, each containing a number of closely separated observations, as illustrated in Figure~\ref{fig:Epochobs} and summarized in Table \ref{tab:epoch_info} (see also Table 1 in \citealp{2017ApJS..228....2L} for a detailed observation log).  
The cumulative exposure is 1.0, 1.0, 2.0 and 3.0 Ms for Epochs 1, 2, 3 and 4, respectively.
The total observational duration, i.e., time between the start of the first observation and the end of the last observation, ranges from 45 days in Epoch 2 to 715 days in Epoch 4. 
Our primary interest below is to search for QPO signals significant over each of these four epochs (Section~\ref{sec:timing}), although we also examine the possible existence of transient QPOs in Section~\ref{subsec:transient}.

We downloaded and uniformly reprocessed the archival data with CIAO v4.13 and CALDB v4.9.4, following the standard procedurex.
The observation with the longest exposure, i.e., ObsID 16462, served as the reference frame when aligning the relative astrometry among the individual observations, which was done by the CIAO tool \emph{reproject\_aspect}.
After obtaining the level-2 event file for each observation, we corrected the photon arrival time to the Solar System barycenter (i.e., Temps Dynamique Barycentrique time) by the CIAO tool \emph{axbary}. 
An exposure map and a point-spread function (PSF) map with 90\% enclosed count fraction (ECF) were generated for each observation, for the photon energy range of 0.5--8 keV.
Both the exposure map and PSF map were weighted by a fiducial spectrum, which is an absorbed power-law with column density $N_{\rm H}=10^{20}\rm~cm^{-2}$ and photon-index $\rm \Gamma=1.4$.
Having examined the light curve of each observation and found that the instrumental background was quiescent for the vast majority of time intervals, we preserved all the science exposures for the subsequent timing analysis, taking the advantage of uninterrupted light curves within each observation.
 
\begin{figure*}
\centering
\includegraphics[scale=0.55]{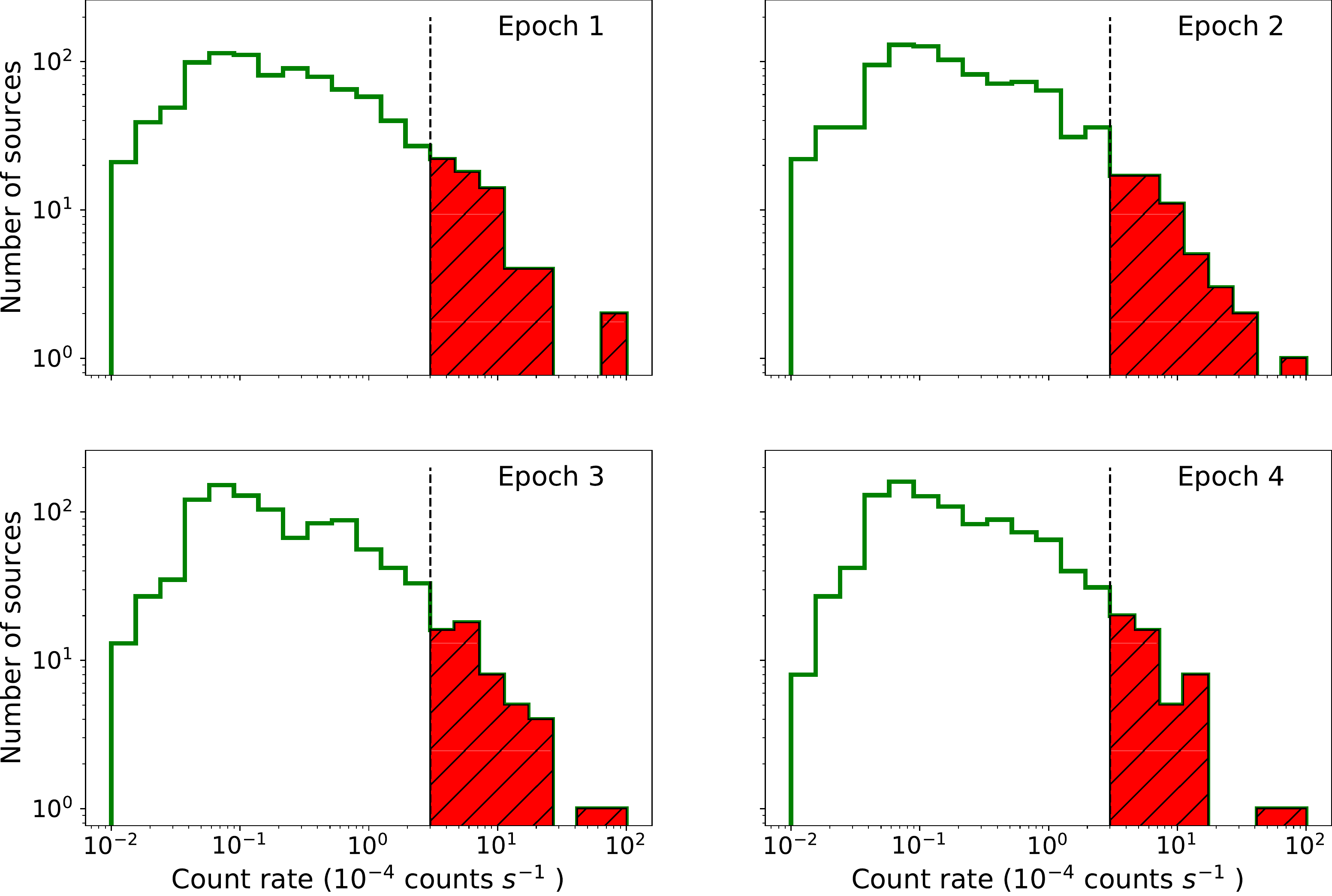}	
\caption{Histograms of the 0.5--8 keV net count rate of the CDF-S sources in the four epochs. 
The red shaded area highlights sources with a net count rate above $3 \times10^{-4}$ cts~s$^{-1}$, a threshold applied in the simulated light curves (Section~\ref{sec:simulation}).}
\label{fig:src_hist}
\end{figure*}

The targets of interest are adopted from the catalog of \citet{2017ApJS..228....2L}, which includes 1055 independent point-like sources.
It is noteworthy that a dozen of CDF-S sources are classified as foreground stars \citep{2017ApJS..228....2L}. 
We include these sources in our analysis for completeness.
For a given source, a light curve was constructed by extracting the 0.5--8 keV events from within the 90\% enclosed count radius (ECR) centering on the given source centroid, while the corresponding background counts were extracted from a concentric annulus with inner-to-outer radii of 2-–4 times the 90\% ECR, excluding pixels falling within two times the 90\% ECR of neighbouring sources, if any. 
Figure~\ref{fig:src_hist} displays histograms of source net count rate in the the four epochs.


\begin{table*}
\centering
\caption{Four Epochs of {\it Chandra} Deep Field-South Observations} \label{tab:epoch_info}
\centering
\begin{tabular}{lcccc}
\hline
\hline
ObsID (in chronicle order) & Start Date & End Date & Total Exposure &  Epoch (duration)\\ 
\hline
581, 1431, 441, 582, 2406, 2405 
\\2312, 1672, 2409, 2313, 2239 & 1999 Oct 14 20:39 & 2000 Dec 23 17:28 & 1.02 Ms & 1 (436 days) \\
\hline
8591, 9593, 9718, 8593, 8597, 8595 \\
8592, 8596, 9575, 9578, 8594, 9596 & 2007 Sep 20 05:26 & 2007 Nov 04 04:11 & 1.01 Ms & 2 (45 days) \\
\hline
12043, 12123, 12044, 12128, 12045 \\
12129, 12135, 12046, 12047, 12137 \\
12138, 12055, 12213, 12048, 12049 \\
12050, 12222, 12219, 12051, 12218 \\
12223, 12052, 12220, 12053, 12054 \\
12230, 12231, 12227, 12233, 12232 \\
12234 & 2010 Mar 18 01:39 & 2010 Jul 22 19:57 & 2.05 Ms & 3 (65 days) \\
\hline

16183, 16180, 16456, 16641, 16457 \\
16644, 16463, 17417, 17416, 16454 \\
16176, 16175, 16178, 16177, 16620 \\
16462, 17535, 17542, 16184, 16182 \\
16181, 17546, 16186, 16187, 16188 \\
16450, 16190, 16189, 17556, 16179 \\
17573, 17633, 17634, 16453, 16451 \\
16461, 16191, 16460, 16459, 17552 \\
16455, 16458, 17677, 1809, 18719 \\
16452, 18730, 16185 & 2014 Jun 09 15:42 & 2016 Mar 24 09:19 & 3.06 Ms  & 4 (715 days) \\
\hline
\end{tabular}
\end{table*}

\section{QPO searching}
\label{sec:timing}
\subsection{Lomb-Scargle Periodogram}
\label{sec:LS}
Previous X-ray detection of AGN QPOs were mostly based on analysis of power spectral density (PSD) on a continuous exposure \citep{2008Natur.455..369G,2013ApJ...776L..10L,2015MNRAS.449..467A,2016ApJ...819L..19P}.
Due to the observing gaps of the CDF-S, we apply the generalized Lomb-Scargle (hereafter LS) periodogram \citep{1976Ap&SS..39..447L,1982ApJ...263..835S}, with a normalization of sample variance following \citet{2009A&A...496..577Z}, to search for QPO signals.
The advantage of the LS periodogram lies in its relatively high operation speed and adequate tolerance for observational gaps. Moreover, the LS periodogram has proved successful in the detection of a few AGN QPOs based on single X-ray observations (e.g., \citealp{2017ApJ...849....9Z,2018A&A...616L...6G,2020A&A...644L...9S}).

 
To construct the LS periodogram, one first defines the frequency grid. 
In principle, the minimum and maximum frequencies should be $1/T$ and $1/(2T_{\rm bin})$, respectively, where $T$ is the total observational duration and $T_{\rm bin}$ is the length of time bins.
A natural time bin is imposed by the ACIS frame time of 3.2 s. However, due to the intrinsic faintness of most CDF-S sources, we instead adopt $T_{\rm bin}=100$ s, which helps suppress the Poisson noise without losing sensitivity to AGN QPOs that empirically exhibit a typical period of hours. 
We further restrict the period searching range between 200 s and 20 ks, since most currently known AGN QPOs exhibit a period shorter than 20 ks, with the exception of XMMU J134736+173403 (85.7 ks; \citealp{2018MNRAS.477.3178C}) as well as two QPEs (32.4 ks in GSN 069, \citealp{2019Natur.573..381M}; 66.6 ks in eRO-QPE1, \citealp{2021Natur.592..704A}).
The size of the frequency grid is taken to be $\Delta \nu= 1/(n_0T)$, where $n_0$ is known as the oversampling factor and chosen to be 5 in our analysis. 

For a given source in a given epoch, a periodogram is thus generated based on the background-subtracted light curve.
We note that a source may not have positive net counts in all four epochs (due either to intrinsic variability or to falling outside the FoV), in which case a periodogram is invalid.
In total we have 3688 valid periodograms from the 1055 independent sources.
The frequency ($\nu$) at which the peak of the LS periodogram is found then indicates a candidate (quasi-) periodic signal.


While the value of the normalized LS periodogram provides some measure of the significance of the periodic signal, it depends on the number of data points and the signal-to-noise ratio, according to the simulations of \citet{2018ApJS..236...16V}. 
This hinders the simple use of the normalized LS periodogram for a direct comparison among a large set of sources.
Therefore, we employ the false alarm probability (FAP), defined below, to assess the significance of a candidate QPO identified from the LS periodogram.  
We note that the FAP essentially measures the probability of a tentative periodic signal arising from pure statistical noise, rather than the probability of a detection of periodic signal.

The traditional approach to quantify the FAP, as proposed by \citet{1982ApJ...263..835S}, is only applicable to data of pure Gaussian noise. 
Another potentially robust way of estimating FAP is through bootstrapping, which not only makes little assumption on the underlying signal, but also can fully account for the observing window. However, the bootstrap method takes a prohibitive computational time even for a single periodogram, making it impractical for a systematic search like our present case.
Hence, we apply the analytic approach proposed by \citet{2008MNRAS.385.1279B} to estimate the FAP, which is based on the theory of extreme values in stochastic processes (see Appendix \ref{sec:FAP} for details).
This so-called \textsl{Baluev method} is known to provide a good approximation (albeit with a slight overestimation compared to the FAP predicted by bootstrapping, see \citealp{2018ApJS..236...16V}), making it particularly suitable for handling sizable data with a highly irregular observing cadence, at only a moderate computational cost.
The FAP quoted in the following refers to that derived from the Baluev method.

\begin{figure*}
\centering
\includegraphics[scale=0.55]{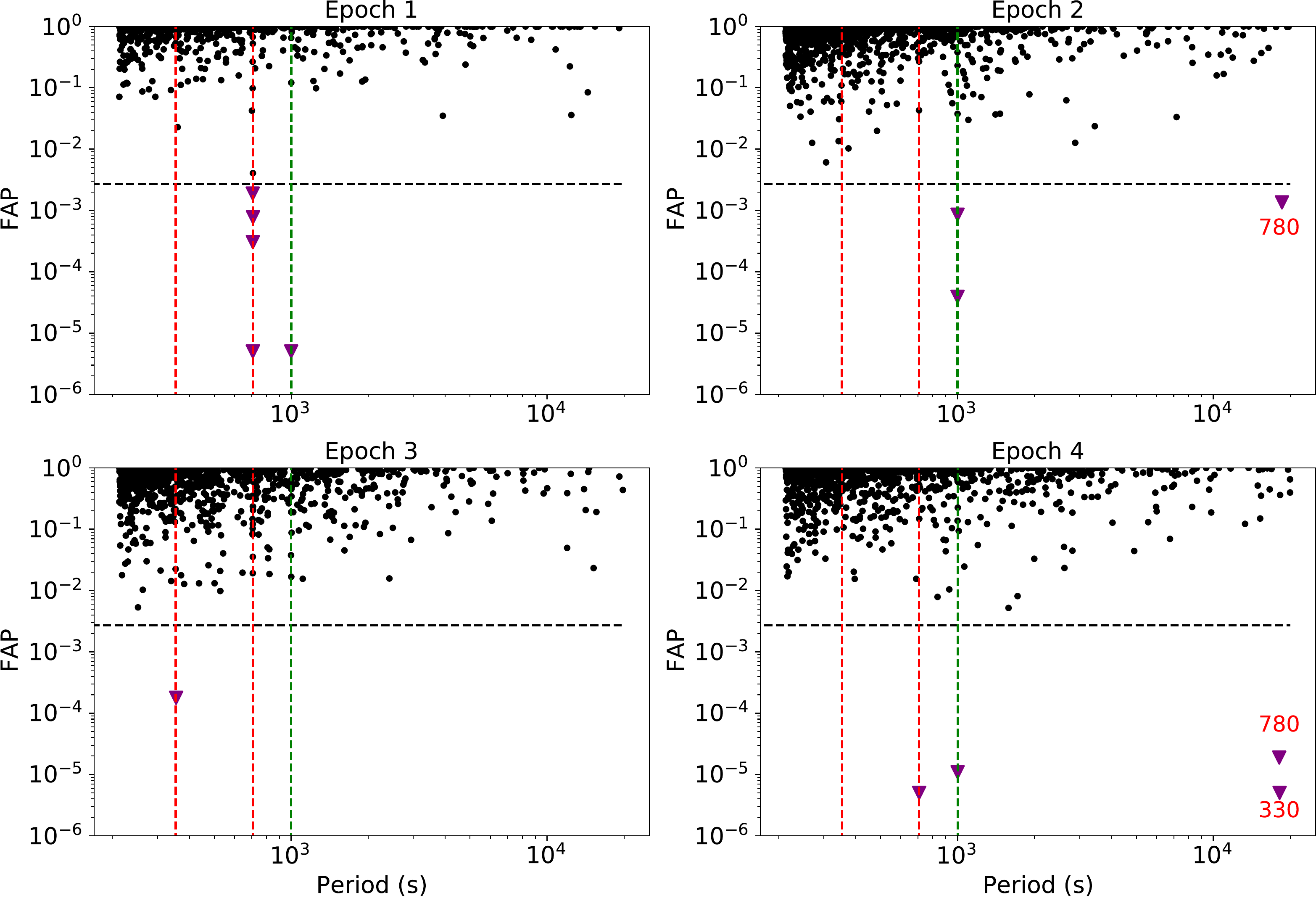}

\caption{The false alarm probability (FAP) of the most probable period identified by the Lomb-Scargle periodogram for the 1055 CDF-S sources in each of the four epochs. 
The black horizontal dashed line marks a threshold of FAP=0.0027, below which a periodic signal is highly unlikely arising from pure statistical fluctuations. 
Tentative signals with FAP below this threshold are shown by purple triangles, while the remaining are represented by black dots. 
The red and green vertical dashed lines mark the periods of telescope dithering at 707 s and 1000 s and their harmonics, which are coincident with most of the tentative signals. Signals associated with two sources of strong red noise, XID 330 and 780, are highlighted. 
}

\label{fig:LSres}
\end{figure*}

\begin{figure*}
\centering
\includegraphics[angle =0, width = 0.49\textwidth]{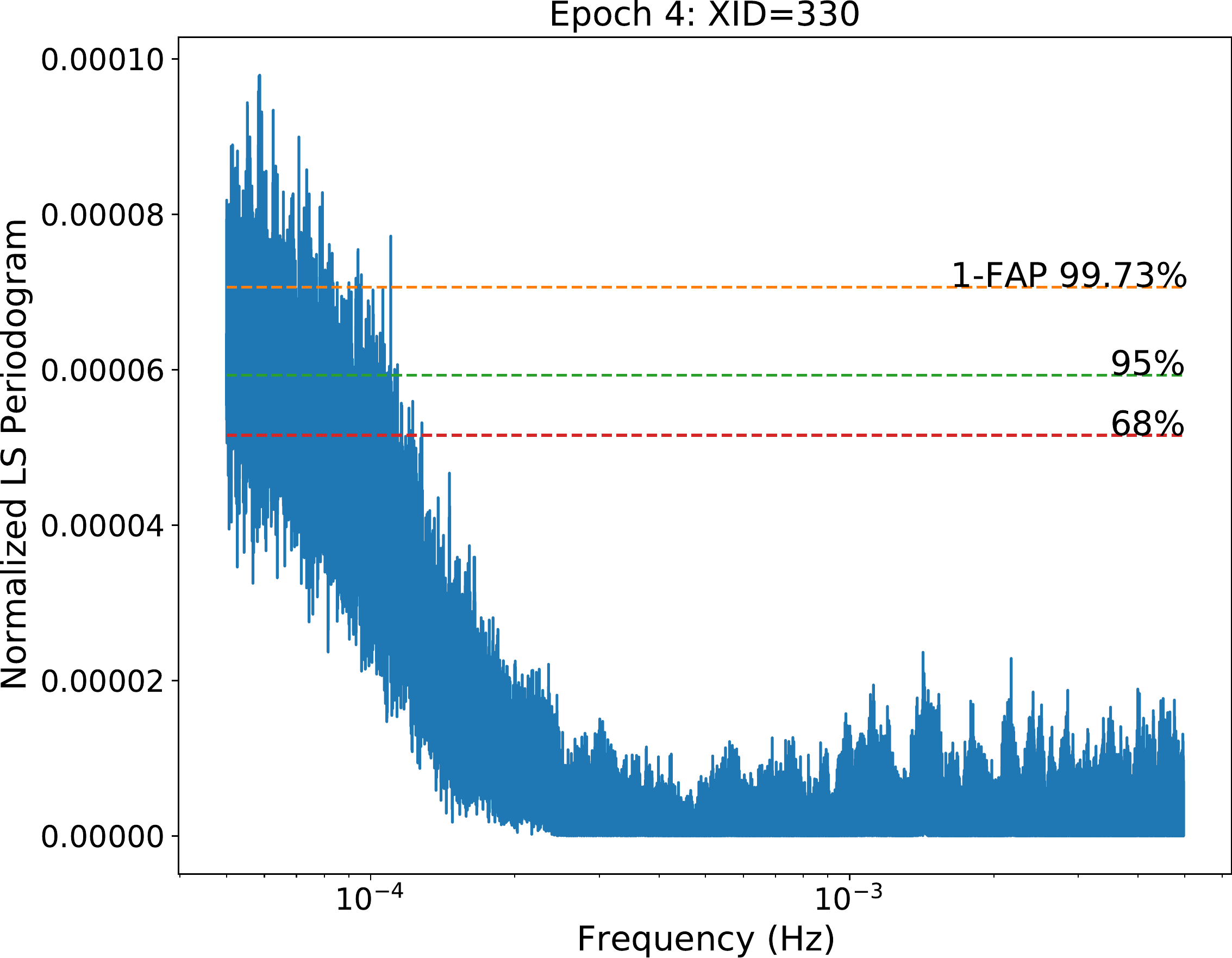}
\hfill
\includegraphics[angle =0, width = 0.49\textwidth]{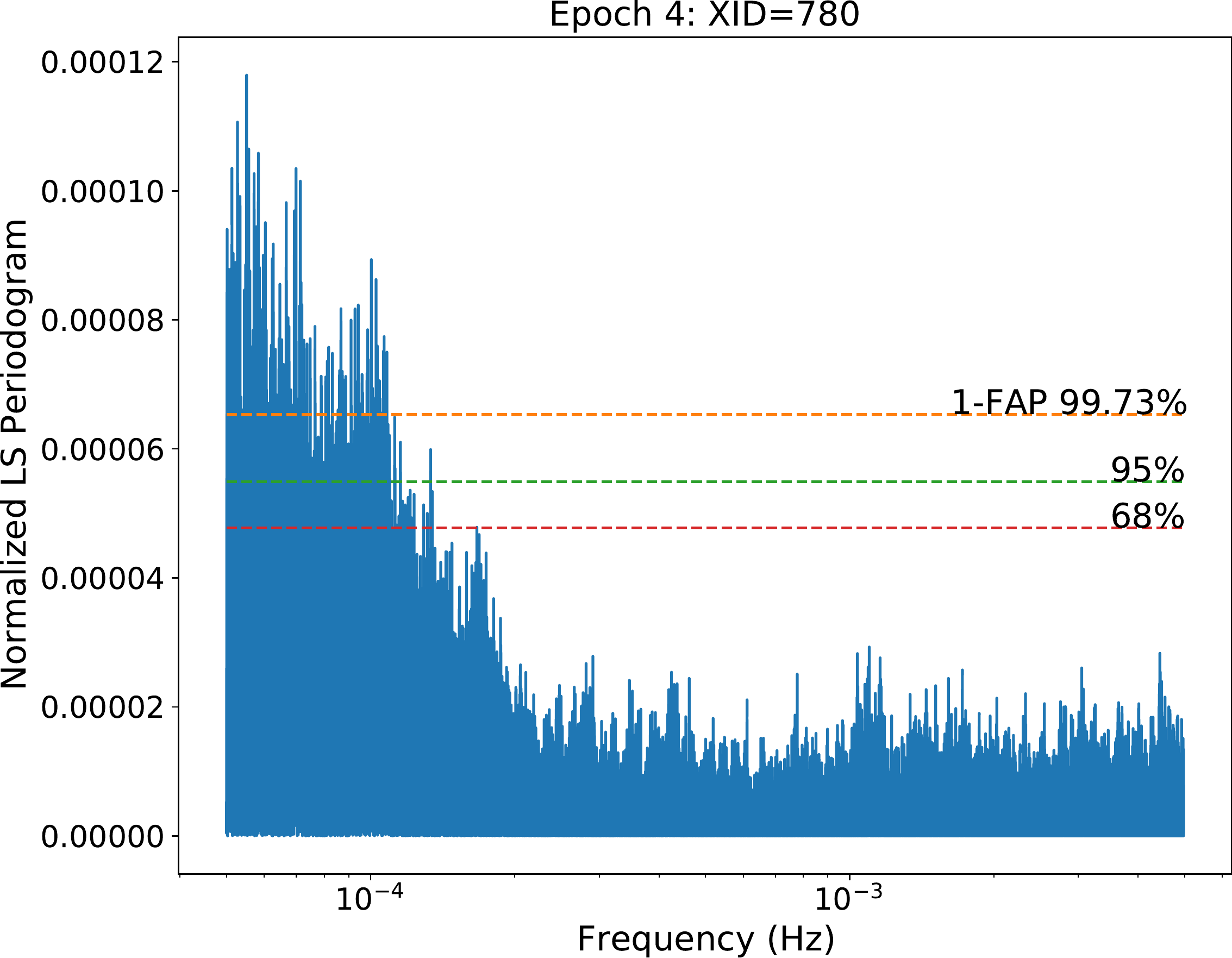}
\caption{The normalized LS periodogram for two sources showing extreme aperiodic flux variations in their light curve, in which case the red noise can cause artificial signals. {\it Left}: XID 330; {\it Right}: XID 780. The orange, green and red dotted lines denote the false alarm probability level of 1--99.73\%, 1--95\%, 1--68\%, respectively.
}
\label{fig:2srcLSP}
\end{figure*}

The period ($P=1/\nu$) at which the highest LS power of each source is found is plotted against the corresponding FAP in Figure \ref{fig:LSres}.
Overall, twelve tentative periodic signals are detected with FAP $< 1-99.73\%=0.27\%$, a threshold below which the signal is considered real. 
We note that the significance of the signal could be even higher if estimated by the bootstrap method.
Among the twelve signals, ten are apparently caused by the dithering motion of the {\it Chandra} telescope, because these are found at a period of 707 s or 1000 s, as well as their harmonics, as labeled by the vertical dashed lines in Figure~\ref{fig:LSres}.
These ten signals are all associated with sources located near the chip gaps of the I-array, such that the dithering motion causes an artificial periodic fluctuation of the source count rate.
The remaining two signals are found in sources XID 330 and XID 780 (as named in the catalog of \citealp{2017ApJS..228....2L}).
XID 330 was reported as an X-ray transient by \citet{2017ApJ...849..127Z}. We find that this source exhibited an outburst in ObsID 16453 belonging to Epoch 4, resulting in severe red noise that fools the LS periodogram with a fake periodic signal at 17043 s in this epoch (illustrated in the left panel of Figure \ref{fig:2srcLSP}).
XID 780, a source classified as a foreground star due to it high proper-motion \citep{2017ApJ...849..127Z}, exhibits a period of 19500 s in both Epoch 2 and Epoch 4. 
A close examination of its light curve shows that this source also exhibited outbursts in ObsIDs 16177 and 16450 (both belonging to Epoch 4) as well as strong variability (ableit with lower amplitudes) in Epochs 1 and 2.
Thus a fake periodic signal could be alarmed with a small FAP, as illustrated in the right panel of Figure \ref{fig:2srcLSP}.

To conclude, the LS periodogram finds no significant {\it genuine} periodic signals among the CDF-S sources.

\subsection{Gregory-Loredo Algorithm}
In addition to the LS periodogram, we employ the Gregory-Loredo (hereafter GL) algorithm \citep{1992ApJ...398..146G} for period searching. 
Essentially a phase-folding method, the GL algorithm excels in finding periodic signals from X-ray data, which is often subject to a moderate number of photon events and/or an irregular observing cadence \citep{2020MNRAS.498.3513B}.
Like most phase-folding methods, the GL algorithm may not be ideal for quasi-periodic signals, since QPOs are often subject to shifting phases or varying amplitudes. 
Nevertheless, we consider the GL algorithm a good complement to the LS periodogram for a blind search of periodic signals from the CDF-S sources. 
Moreover, results from such an exercise would provide a useful empirical estimate of periodic X-ray signals randomly arising from the cosmic background, which is currently unconstrained but may affect the study of periodic X-ray sources in Galactic fields (e.g., \citealp{2020MNRAS.498.3513B}). 


Unlike the LS periodogram, the GL algorithm manipulates photon arrival times rather than the binned light curve. Thus the arrival times of 0.5--8 keV events of a given source are taken as direct input and there is no need for background subtraction (that is, absorbed into the constant part of the source light curve). 
We also divide the 102 observations into the same four epochs to be consistent with Section \ref{sec:LS}. 

Similarly, we obtain 3688 valid time series for the 1055 sources, which are then analyzed with the GL algorithm following the procedures detailed in \citet{2020MNRAS.498.3513B}. 
We adopt the same period searching range (200 s to 20 ks) and frequency resolution of $\Delta \nu= 1/n_0T$ as for the LS periodogram. 

\begin{figure*}
\centering
\includegraphics[scale=0.55]{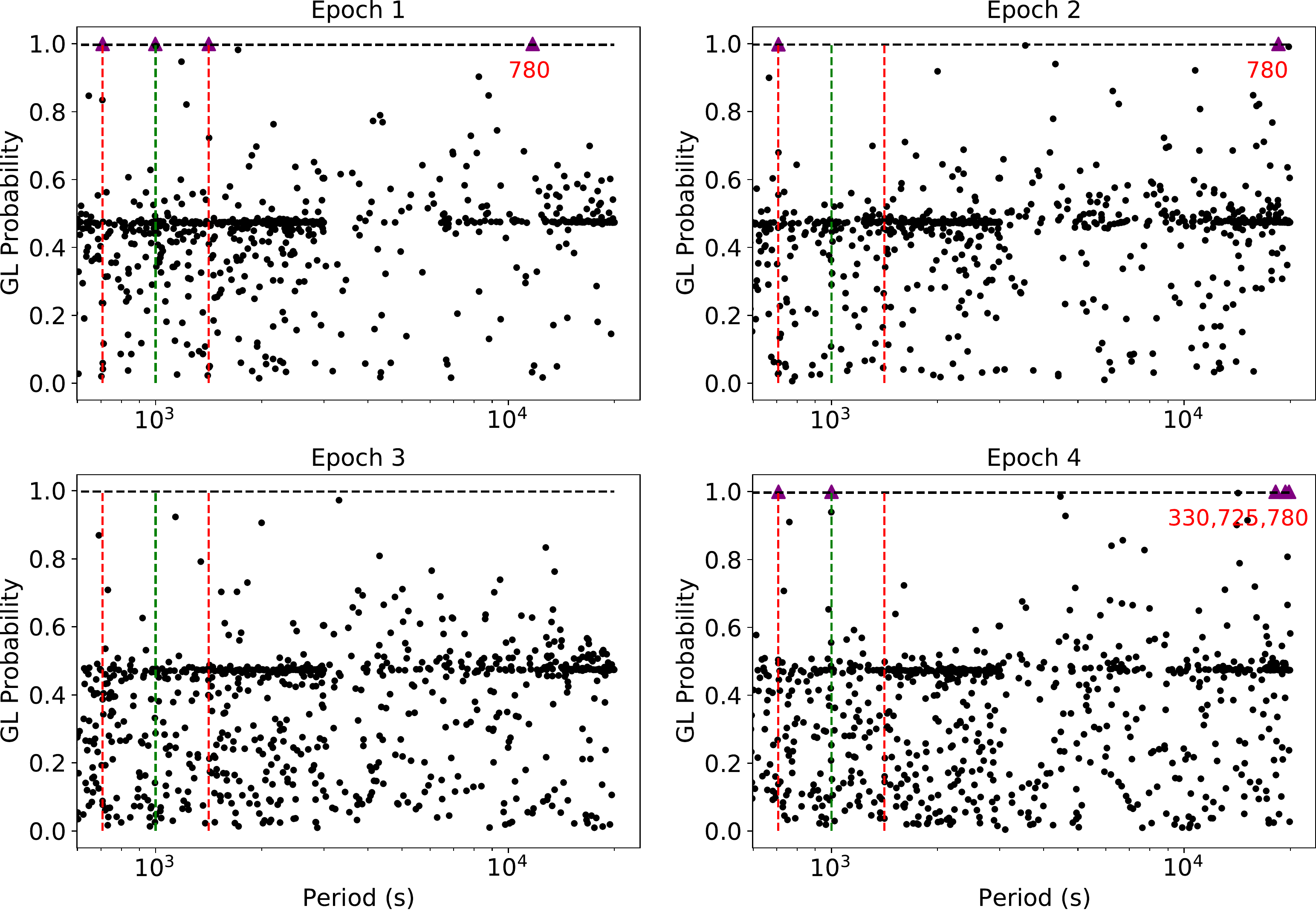}

\caption{The probability of a detected period by the Gregory-Loredo Algorithm for the 1055 CDF-S sources in each of the four epochs. 
The black horizontal dashed line marks a threshold probability of 0.9973. 
Tentative signals exceeding this threshold are shown by purple triangles, while the remaining are represented by black dots. 
The red and green vertical dashed lines mark the periods of telescope dithering at 707 s and 1000 s and their harmonics, which are coincident with most of the tentative signals. Signals associated with three sources of strong red noise, XID 330, 725 and 780, are highlighted. 
}
\label{fig:GLres}
\end{figure*}

Figure \ref{fig:GLres} exhibits the period-searching results for the four epochs, in which the most probable period of each CDF-S source is plotted against the detection probability.
We emphasize that the probability here refers to the probability of a periodic source as defined by the GL algorithm (see a concise introduction and mathematical formulae in the Appendix of \citealp{2020MNRAS.498.3513B}).
In all four panels, an obvious concentration of sources is seen around a probability of 0.5, 
which in fact comprises of low-count sources.  
Mathematically, as predicted by Eqn. A19 of \citet{2020MNRAS.498.3513B}, the GL probability approaches a value of 0.5 for source counts below $\sim$50, since this is when the phase distribution becomes dominated by random noise. 

Adopting a threshold of 99.73\% for a significant periodic signal and excluding a few false detections corresponding to the periods of telescope dithering (marked by the vertical dashed lines in Figure \ref{fig:GLres}), only three sources are picked up by the GL algorithm. 
Two of the three sources (XID 330 and XID 780) have already been identified by the LS periodogram due to their strong red noise. 
The third source, XID 725, was also reported as a transient by \citet{2017ApJ...849..127Z} (see also \citealp{2017MNRAS.467.4841B} for more discussions about the properties of XID 725 and XID 330), which exhibited outbursts in ObsID 16453 and 16454 (both belonging to Epoch 4).
It is known that the GL algorithm may also be fooled by strong aperiodic variability \citep{2020MNRAS.498.3513B}. 
Thus these three ``periodic'' signals are almost certainly false detections.
This is supported by a re-analysis of a subset of the light curve after excluding the time intervals of strong variability; the tentative period identified based on the full time intervals in fact cannot be recovered in either subset of the three sources. 

While it is computationally straightforward to feed the GL algorithm with a single time series combining all four epochs, we simply note in passing that no extra periodic signal is found in this case, even when the period searching range is extended up to 100 ks.
We conclude that the GL algorithm also finds no significant genuine periodic signals among the CDF-S sources.

Lastly, we note that a small fraction of the 1055 sources actually could be more significant in a sub-band (e.g., 0.5--2 keV) than in the full 0.5--8 keV band \citep{2017ApJS..228....2L}. We have also examined sub-band light curves but found no extra significant QPO signals, either from LS periodogram or GL algorithm.

\subsection{Potential caveat due to red noise}
\begin{figure}
\centering
\includegraphics[scale=0.56]{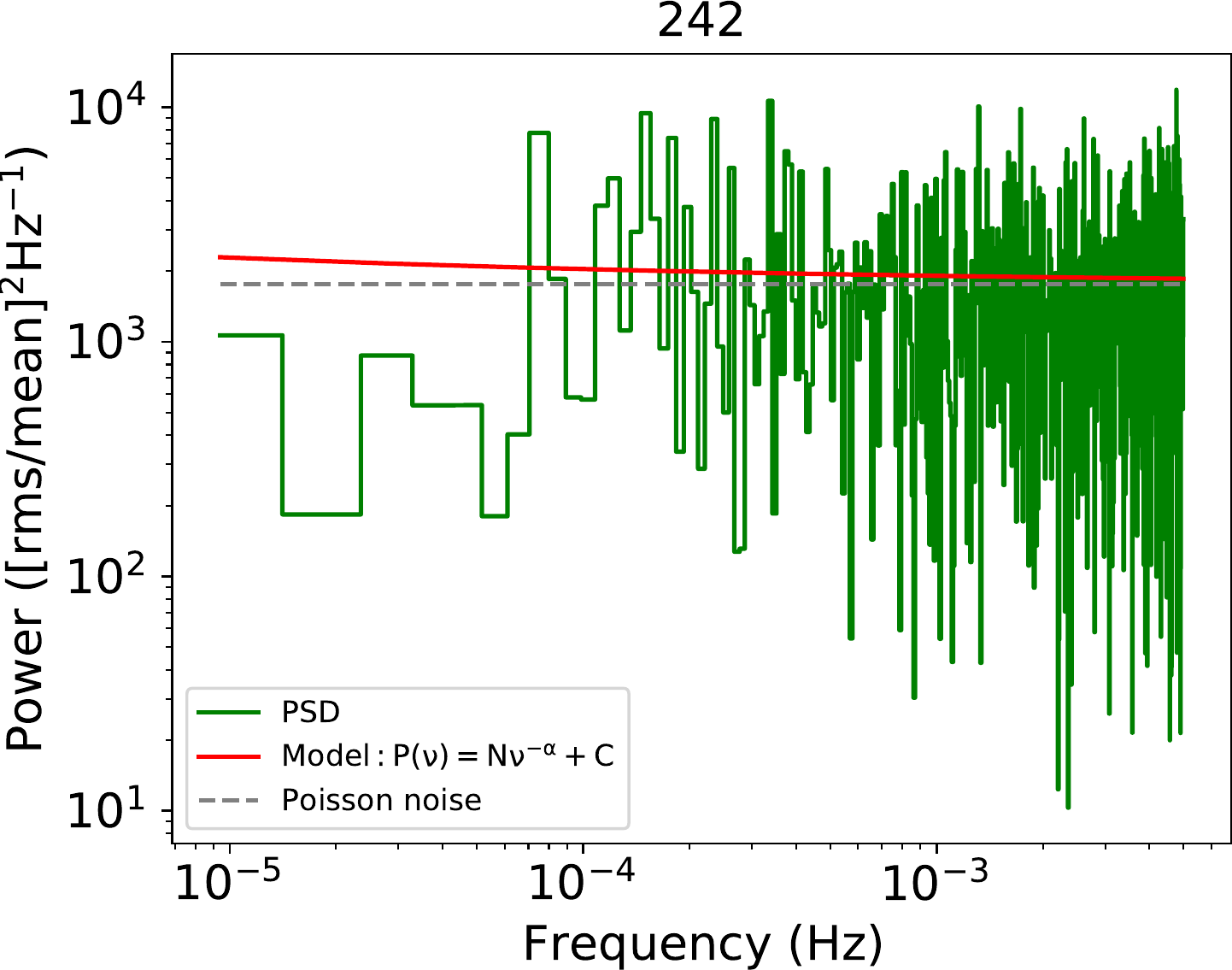}

\caption{The power spectral distribution (green curve), normalized to $\rm (rms/mean)^2$ per Hz, of a representative bright source XID 242 in ObsID 16188.
The best-fit model and the expected Poisson noise level are represented by the red solid line and grey-dashed line, respective.}
\label{fig:242psd}
\end{figure}

The above filtering of fake signals due to strong red noise raises the concern of whether true periodic signals have been masked by red noise, given the fact that neither the LS periodogram nor the GL algorithm is designed to optimally handle red noise. 
A few methods \citep{1996ApJ...468..369I,2005A&A...431..391V,2010MNRAS.402..307V}, mostly based on Fourier analysis, have been proposed in the literature to reveal periodic signals out of red noise, which is especially relevant to bright AGNs with flickering variability. 
However, in the vast majority of CDF-S sources, Poisson noise actually dominates the source power spectrum due to a low-to-moderate count rate, even in the presence of intrinsic red noise. 
This is illustrated in Figure \ref{fig:242psd}, which presents the normalized power spectrum of XID 242, one of the brightest sources in CDF-S (with a net count rate $\rm \sim {10}^{-3}~cts~s^{-1}$), as observed in a single observation, ObsID 16188. 
A power-law plus constant model, $P(\nu)=N\nu^{-\alpha}+C$, is adopted as the PSD template (see more discussions in Section~\ref{sec:simulation}), where $N$ is the normalization factor, $\alpha$ is the slope, and $C$ is a constant representing the photon counting induced Poisson noise. We follow the method introduced by \citet{2010MNRAS.402..307V} to derive the model parameters. It can be seen that the best-fit model (red solid line) is totally dominated by the constant part, i.e., the Poisson noise, which is represented by the grey-dashed line. 
For fainter sources, the lower count rate brings to even higher Poisson noise. 
Therefore, we can conclude that red noise is unlikely to mask a true periodic signal in the periodogram of the CDF-S sources.

\section{Simulation}
\label{sec:simulation}
The non-detection of periodic signals from the $\sim$ 1000 CDF-S sources may be understood as one of the two following reasons: (i) an intrinsically low occurrence rate of AGN QPOs, and/or (ii) a low sensitivity of the CDF-S data in detecting QPO signals. 
In this section, we employ simulated AGN light curves to evaluate the second possibility. The results may also be used to examine the first possibility, as further discussed in Section~\ref{sec:discussion}.

\subsection{The power spectrum model}
\label{subsec:psd}
 Following the method proposed by \citet{1995A&A...300..707T}, we generate mock data that exhibit typical power spectrum of AGNs. 
We adopt a bending power-law function to describe the PSD of a typical AGN,

\begin{equation}
	P_{\rm b}(\nu)=N\nu^{-1}{(1+(\frac{\nu}{\nu_{\rm b}})^{\alpha-1})^{-1}},
\label{eqn:bendpl}
\end{equation}
where $\nu_{\rm b}$ is the bend frequency, $\alpha$ is the spectral index above $\nu_{\rm b}$, and $N$ is the normalization. 
Such a PSD shape is found to be common in the AGN sample of \citet{2012A&A...544A..80G} and is understood as the manifestation of flickering noise in the time domain, which may be caused by thermal viscous instability or a variable mass accretion rate \citep{2017MNRAS.470.3027K}.

The PSD of a QPO can be described by a Lorentzian function, 
\begin{equation}
	P_{\rm L}(\nu)=\frac{R^2Q\nu_0/\pi}{{\nu_0}^2+Q^2(\nu-\nu_0)^2},
\label{eqn:lorentz}
\end{equation}
where $\nu_0$ is the central frequency of the Lorentzian, $Q =\nu_0 /\delta \nu$ is the quality factor (2$\delta \nu$ is the full width at half height), and $R$ is a normalization factor. In the case of high $Q$ values, $R$ would be approximately equal to the fractional rms amplitude of the QPO.

Then the intrinsic PSD of an AGN with an QPO can be described by the sum of Eqn.~\ref{eqn:bendpl} and Eqn.~\ref {eqn:lorentz}, 
\begin{equation}
	P(\nu)=N\nu^{-1}{(1+(\frac{\nu}{\nu_{\rm b}})^{\alpha-1})^{-1}}+\frac{R^2Q\nu_0/\pi}{{\nu_0}^2+Q^2(\nu-\nu_0)^2}.
\label{eqn:bendplgauss}
\end{equation}

Due to the limited counts of the CDF-S sources, it is impractical to obtain their individual intrinsic PSDs. 
For the sake of simulating source light curves, we adopt representative parameters for a modeled PSD, as justified below.

The bend frequency, $\nu_{\rm b}$, is empirically related to both the black hole mass and the accretion rate \citep{2006Natur.444..730M}. The spectral index, $\alpha \approx 3$, gives the asymptotic change of the power-law slope from -1 at the limit of $\nu \ll \nu_{\rm b}$ to -2 for $\nu \gg \nu_{\rm b}$. We find that the exact values of both of these two parameters have little effect on the sensitivity of QPO searching.
Hence we adopt fiducial values of $\nu_{\rm b}=4.3 \times 10^{-4}$ Hz and $\alpha=3.4$, which are from the observed power spectrum of RE J1034+396, the first AGN with a confirmed QPO signal \citep{2010MNRAS.402..307V}. 

On the other hand, the normalization of the bending power-law, $N$, which represents the amplitude of the intrinsic noise, has a strong effect on the QPO searching. According to the observational results by \citet{2012A&A...542A..83P}, $N$ increases with decreasing Eddington ratio in the X-ray band, $\lambda_{\rm Edd,X} =L_{\rm X}/L_{\rm Edd}$ (i.e., ratio of the X-ray luminosity to the Eddington luminosity),
\begin{equation}
N=2 \nu_{\rm b} P_{\rm b}(\nu=\nu_{\rm b}) \approx 3\times 10^{-3}\lambda_{\rm Edd,X}^{-0.8}.
\label{eqn:normalization}
\end{equation}
For classical AGNs accreting near the Eddington limit, $\lambda_{\rm Edd,X}$ has a characteristic value of $\sim$ 10\%, suggesting $N \sim 0.02$.
We also consider a lower value of $N=2.3\times 10^{-3}$, which is again derived from the power spectrum of RE J1034+396 \citep{2010MNRAS.402..307V}. 


Among the three parameters of the Lorentzian function,
the normalization factor $R$ is the most crucial one for the QPO detectability, since it determines the amplitude of the signal. 
$R$ ranges from 5\% to 15\% in the currently known cases of NLS1s. 
In the two cases of Seyfert 2s, $R$ is as high as $25-50\%$ \citep{2013ApJ...776L..10L}.
The quality factor, $Q$, which determines the width of the QPO, is often not well constrained due to insufficient frequency resolution, i.e., the reciprocal of observation time. 
We adopt $Q$ = 15 as fiducial, which is again from the observed value of RE J1034+396 and not atypical among known AGN QPOs. 
The frequency $\nu_0$ sets the QPO period, which for Seyfert 1 galaxies falls between 1 hour to 2 hour, while for the two cases of Seyfert 2 is as long as 23.8 hour \citep{2018MNRAS.477.3178C}. 
We focus on periods between 1--2 hour, noting that in general the detection efficiency of QPO signals is higher for higher periods, other conditions being equal (see Section~\ref{subsec:DE}). 

Given the above considerations, we define two PSD models both including a QPO. 
{\it Model A} takes $N$=2.3 $\times 10^{-3}$ and $R$=5\%,
while {\it Model B} takes $N$=2 $\times 10^{-2}$ and $R$=15\%.
These two models represent, respectively, the case of low RMS variability but weak QPO signal (RE J1034+396-like) and the case of high RMS variability but strong QPO signal.
The other three parameters,  $\nu_{\rm b}=4.3\times10^{-4}$ Hz, $\alpha=3.4$ and $Q=15$, are fixed for both models.
Lastly, we consider for each model four different values for $\nu_0$: $\rm 1/3600, 1/5400, 1/7200, 1/18000 ~Hz$, corresponding to a period of 1, 1.5, 2, 5 hour, respectively. 
The last value is taken as representative of unusually long QPO periods.

In addition, we examine a {\it no-QPO model}, that is, taking $R$=0 and the remaining parameters identical to those of Model B. This last model provides a consistency check on the false detection rate of QPOs.

\subsection{Producing the light curve}
\label{subsec:simlc}
We apply the algorithm described in \citet{1995A&A...300..707T} to generate X-ray light curves, $\{x(t)\}$, from the modeled PSD (Eqn.~\ref{eqn:bendplgauss}), where $x$ denotes count rate in the time series.
The algorithm in \citet{1995A&A...300..707T} realizes a stochastic process with a zero mean flux. 
Thus a mean count rate must be added to shift the light curve in accordance with a real source.
To mimic the photon counting process with a Poisson noise, we next generate an event list and resample it to a binned light curve.
Our approach is summarized in the following steps:

(1) Generate $\{x_1(t_i)\}$ from a given set of parameters of the PSD, $P(\nu; N, \nu_{\rm b}, \alpha, \nu_0, Q, R)$, and a predefined mean count rate $\overline{x}$, for $t_i$ between $t_0$ and $t_{\rm max}$ according to the start and end times of each real CDF-S observation. For each modeled PSD, we consider $\bar{x}= (3,4,5,6,7,8,9,10) \times10^{-4}$ cts~s$^{-1}$; lower mean count rates have been tested but are not further considered since they would result in a negligible detection efficiency (see Section~\ref{subsec:DE});

(2) Simulate a series of photon arrival time $\{{T_j}\}$ from 
$\{x_1(t_i)\}$ obtained in Step (1), according to a Poisson randomization process;

(3) Produce the output binned light curve $\{x_2(t_i)\}$ from $\{{T_j}\}$ in 100 s time bins;

(4) For each source, repeat the above steps for 1000 times.
%


For a given set of parameters, $102\times1000$ light curves are thus created, which follow the exact observing cadence and are further grouped into four epochs as defined in Section \ref{sec:obs}.
These light curves are then fed to the LS periodogram to search for a periodic signal in the same way as described in Section~\ref{sec:LS}.
We do not further consider the case of the GL algorithm, except for noting that a similar result is anticipated.

\subsection{Detection efficiency} 
\label{subsec:DE}


The period-searching results based on the simulated curves are used to evaluate the detection efficiency (DE) of QPO signals.
We define DE as the percentage of valid detections among the 1000 realizations of a given set of PSD model parameters. 
Here a valid detection of periodic signal should satisfy the conditions of (i) the FAP of the signal peak not exceeding the given threshold (0.27\%, same as in Section~\ref{sec:LS}) and (ii) the peak frequency $\nu_{\rm peak}$ being consistent with the input frequency, $|\nu_{\rm peak}-\nu_0|<\nu_0/(2Q)$. 

\begin{figure*}
\centering
\includegraphics[scale=0.55]{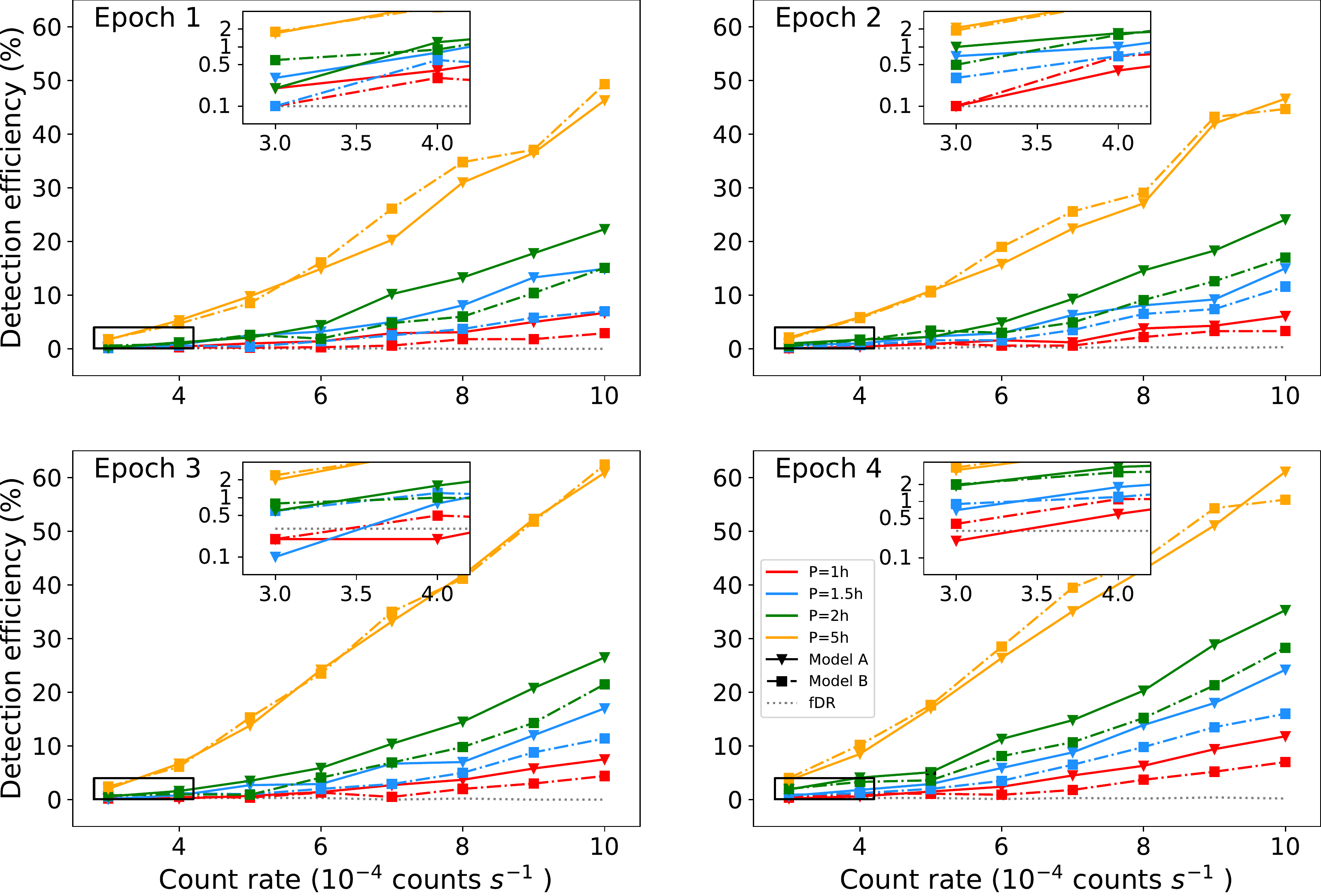}
\caption{QPO detection efficiency of the LS periodogram as a function of mean source count rate and quasi-period, based on simulated light curves. 
The two PSD models are represented by triangles linked by solid lines (Model A) and squares linked by dot-dashed lines (Model B), respectively.
Red, blue and green lines and orange symbols represent a quasi-period of 1, 1.5, 2 and 5 hour, respectively.
The grey dotted line denotes the false detection rate of the no-QPO model, confirming the FAP threshold of 0.27\%. 
The insert in each panel is a zoom-in to better visualize the generally low detection efficiency at the lowest count rates simulated.} 
\label{fig:LSDR}
\end{figure*}

Figure~\ref{fig:LSDR} displays the resultant DE of both Model A and Model B, for different values of the mean count rate and the input QPO frequency. 
The four panels show the case of the four epochs, respectively.

For comparison, the false detection rate (fDR) from the no-QPO model is also plotted, in which a "false detection" is reported when FAP of the highest peak is lower than the given threshold (0.27\%).
fDR is found to be $\lesssim0.3\%$ for all modeled count rates, confirming the correctness of the calculated FAP.

It can be seen that the value of DE ranges from $\lesssim$1\% to $\sim$60\%. 
For both Model A and Model B, DE generally increases with the mean count rate, which is due to the reduction of Poisson noise.
%
According to \citet{2003MNRAS.345.1271V},
if a light curve is composed of binned photon counting signal, the expected Poisson noise level in the rms-normalized periodogram is given by,
\begin{equation}
\label{eqn:noise}
	P_{\rm noise}=\frac{2(\bar{x}+B)}{\bar{x}^2}\frac{\Delta T_{\rm samp}}{\Delta T_{\rm bin}}\approx \frac{2}{\bar{x}}\frac{\Delta T_{\rm samp}}{\Delta T_{\rm bin}},
\end{equation}
where $\Delta T_{\rm samp}$ and $\Delta T_{\rm bin}$ denote the sampling interval (here order of days) and the time bin width (here 100 s), respectively, and $B$ is the background count rate, which is negligible for the majority of the CDF-S sources. 
While not shown in Figure~\ref{fig:LSDR}, we note that DE becomes lower than the threshold of 0.27\% at count rates below $3\times10^{-4}{\rm~cts~s^{-1}}$. 
This applies to the majority of the CDF-S sources (Figure~\ref{fig:src_hist}), which are now understood to fail to exhibit a detectable QPO against the noise. 
 
Figure~\ref{fig:LSDR} reveals another general trend that DE increases with the QPO period. 
This can be attributed to the peak value of the Lorentzian function, $P_{\rm L}(\nu=\nu_0)=R^2 Q/\pi \nu_0$, increasing with the QPO period.  
For the same mean count rate and periods between 1--2 hour, DE is always higher in Model A (triangles linked by solid curves) than in Model B (squares linked by dot-dashed curves). This is primarily due to a lower rms variability in the former, more than compensating for its relatively weak QPO signal. 
Such a trend, however, does not hold for the 5 hour period, which can be explained by the high peak value of the Lorentzian function at this long period, dominating over the rms variability.

\section{Discussion}
\label{sec:discussion}
\subsection{Constraining the occurrence rate of persistent QPOs in the CDF-S}
\label{subsec:rate}
The simulations presented in Section~\ref{sec:simulation} suggest that the sensitivity of the CDF-S data in QPO detection is only moderate. 
This is in accord with the non-detection of QPOs in each of the four epochs (Section~\ref{sec:timing}).
We now take a step forward to assess the intrinsic QPO occurrence rate, under the assumption that all AGNs can possess QPOs that are persistent on a timescale of months to years, i.e., much longer than the period itself. 
Then, the expected number of detectable QPOs follows
\begin{equation}
	N_{\rm QPO,det}=DE\times f_{\rm QPO} \times  N_{\rm sou},
\label{eqn:Ndet}
\end{equation}
where $f_{\rm QPO}$ is the intrinsic fraction of AGNs with a (persistent) QPO and $N_{\rm sou}$ is the total number of sources. 
The twelve classified stars are excluded here since we are only concerned with AGN QPOs.
Simulations in Section~\ref{sec:simulation} have shown that DE varies with QPO period and source mean count rate. Thus the QPO fraction can be expressed as 
\begin{equation}
	f_{\rm QPO} \approx N_{\rm QPO,det}/ (\sum_i{DE(\bar{x}_i,P)\times N_i}), 
\label{eqn:fQPO}	
\end{equation}
assuming that $f_{\rm QPO}$ is independent of the count rate and period (while DE is). 
To evaluate $f_{\rm QPO}$, we divide the CDF-S sources into different bins of mean count rate, from $3\times 10^{-4}$ to $10\times10^{-4}\rm~cts~s^{-1}$ at a step of $10^{-4}$. 
About 5\% of the total sources fall within this range,
while the majority of remaining sources have a lower count rate (Figure~\ref{fig:src_hist}).
These weak sources have a very small DE and their collective contribution in Eqn.~\ref{eqn:fQPO} is negligible.
For the few brightest sources with a mean count rate up to $7\times10^{-3}\rm~cts~s^{-1}$, we use a simple extrapolation to estimate their DE, noting that the exact value has little effect in the result. 
In addition, we average out the dependence of DE on the intrinsic period by using equal weights among the simulation results of $P$ = 1.0, 1.5 and 2.0 hour. 
It is noteworthy that QPOs with longer periods are expected in more massive black holes, the 
detection efficiency of which could be higher, as addressed in Section~\ref{subsec:DE}.
A similar expectation applies to QPOs in type-2 Seyferts, as suggested by the two currently known cases with $P \sim$ 3.8 hour and $\sim$ 23.8 hour. 
Nevertheless, we neglect the contribution from periods longer than 2 hour, noting that lending weight to the DE of long-period signals would simply reduce the resultant $f_{\rm QPO}$, because longer periods give a substantially higher DE (Figure~\ref{fig:LSDR}).

Adopting the simulated DE in Figure~\ref{fig:LSDR}, we have $f_{\rm QPO} <$ 1/6.0, 1/6.0, 1/6.5 and 1/8.8 for QPO Model A of the four epochs. Models B leads to similar constraints, i.e., $f_{\rm QPO} <$ 1/4.6, 1/4.6, 1/5.0 and 1/6.8 for the four epochs.
On average, these values suggest an upper limit of 15\%--20\% for the intrinsic fraction of AGN QPOs that have a similar PSD shape with a handful of currently known QPOs and are persistent for a timescale of months to years.

We note that this might be a loose upper limit, since we have not detected any QPO, leaving the possibility that a much larger population of AGNs is required to survey to find a single signal. 
Indeed, in their search of QPO signals in bright AGNs, \citet{2012A&A...544A..80G} could find only one significant QPO among 104 sources.
While the high count rates of these AGNs (typically 0.1 cts~s$^{-1}$) likely guarantees a high detection efficiency,  
the inhomogeneous XMM-Newton observations and heterogeneous AGN sample, however, prevent a meaningful quantification of the true QPO occurrence rate.    
On the other hand, \citet{2012MNRAS.426.1701B} found a very low detection rate ($0.15\%$) of HFQPOs from 22 Galactic BHBs based on a highly homogeneous analysis of RXTE observations. 
This likely provides a lower limit of the intrinsic occurrence rate of the HFQPOs and may serve as a benchmark for future large survey of AGN QPOs.

\subsection{Transient QPO candidates}
\label{subsec:transient}
So far we have focused on QPO signals significant over a duration of tens to hundreds of days.
This leaves the possibility that short-lived, or transient, QPOs are missed in our above search. 
In fact, a number of known AGN QPOs are transient QPOs, i.e., they were only detected over single X-ray observations or even over a fraction of the observation (e.g., \citealp{2016ApJ...819L..19P,2017ApJ...849....9Z,2020A&A...644L...9S}), such that the duration of activity is not much longer than the period itself.

To look for transient QPOs, we apply the LS periodogram to the short-term light curves of all 1055 sources.
Here a short-term light curve is defined as one that contains any number of successive observations within a certain epoch. 
For instance, from the 12 individual observations of Epoch 2, one may construct for a given source 11 unique light curves that contain two neighboring observations, regardless of the length of the gap.   
The total number of unique light curves within an epoch of $N_{\rm obs}$ observations is thus given by $ \sum_{i=1}^{i=N_{\rm obs}} (N_{\rm obs}-i+1) = \frac{1}{2}N_{\rm obs}(N_{\rm obs}+1)$. 

\begin{figure*}
\centering
\includegraphics[angle =0, width = 0.32\textwidth]{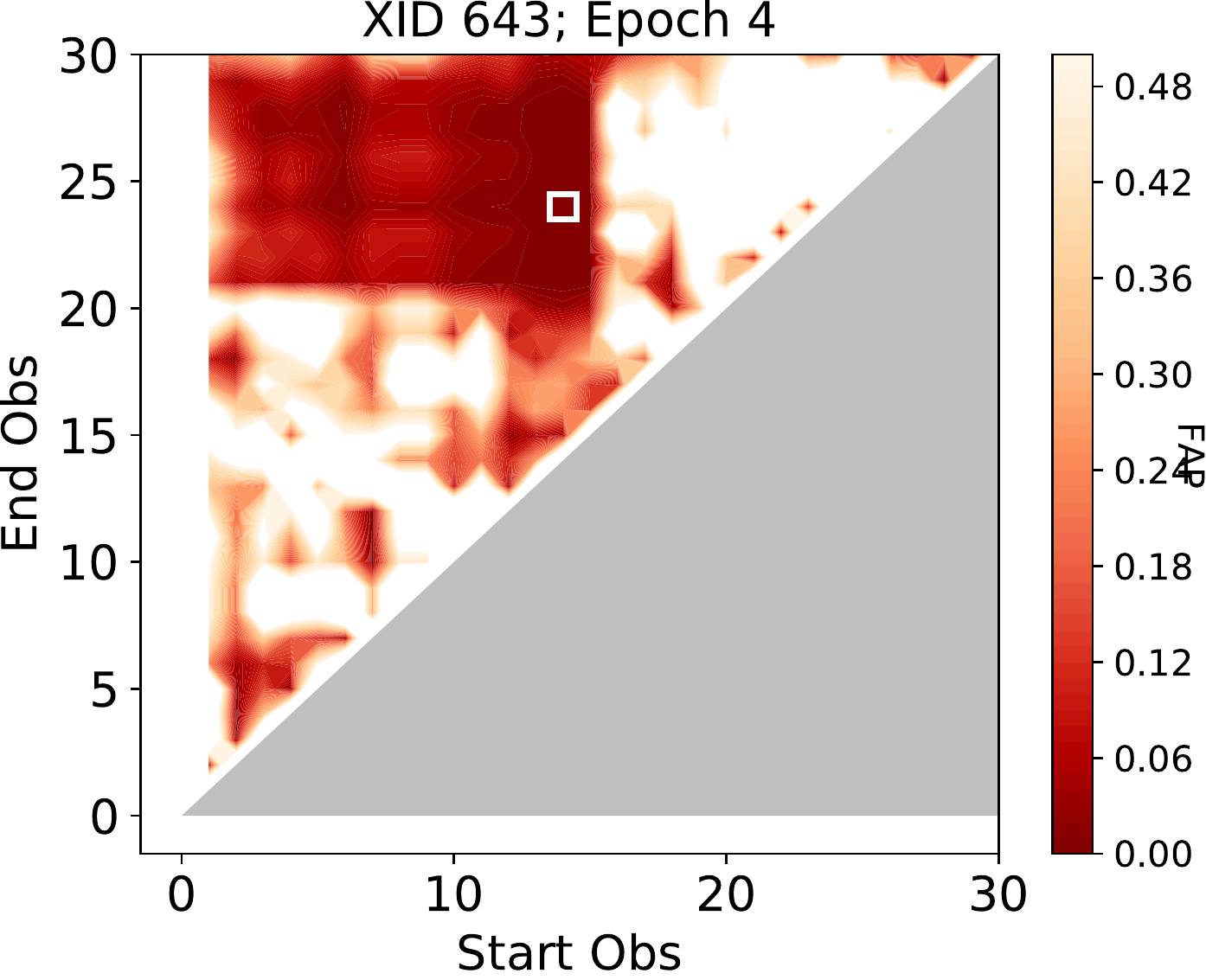}
\hfill
\includegraphics[angle =0, width = 0.32\textwidth]{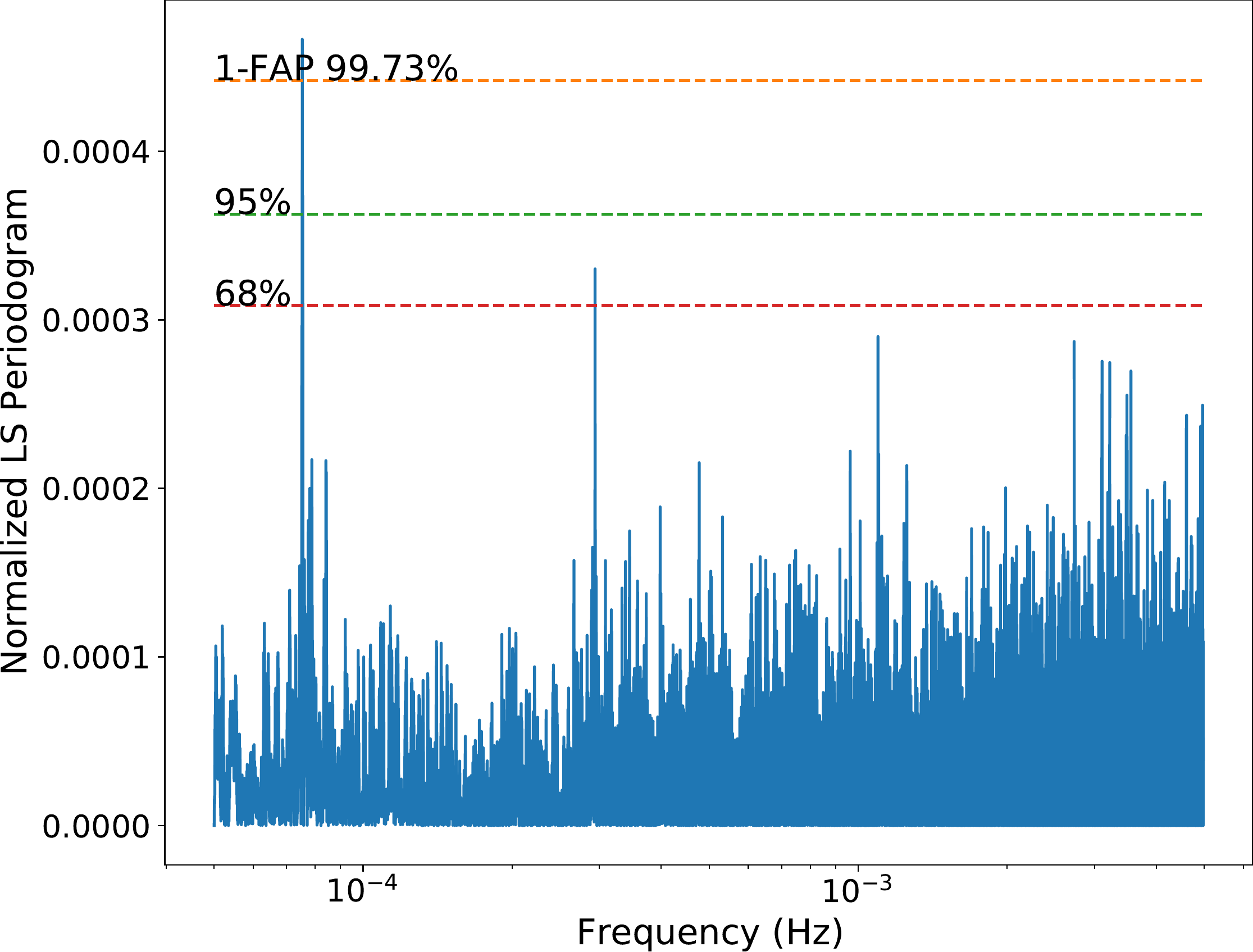}
\hfill
\includegraphics[angle =0, width = 0.32\textwidth]{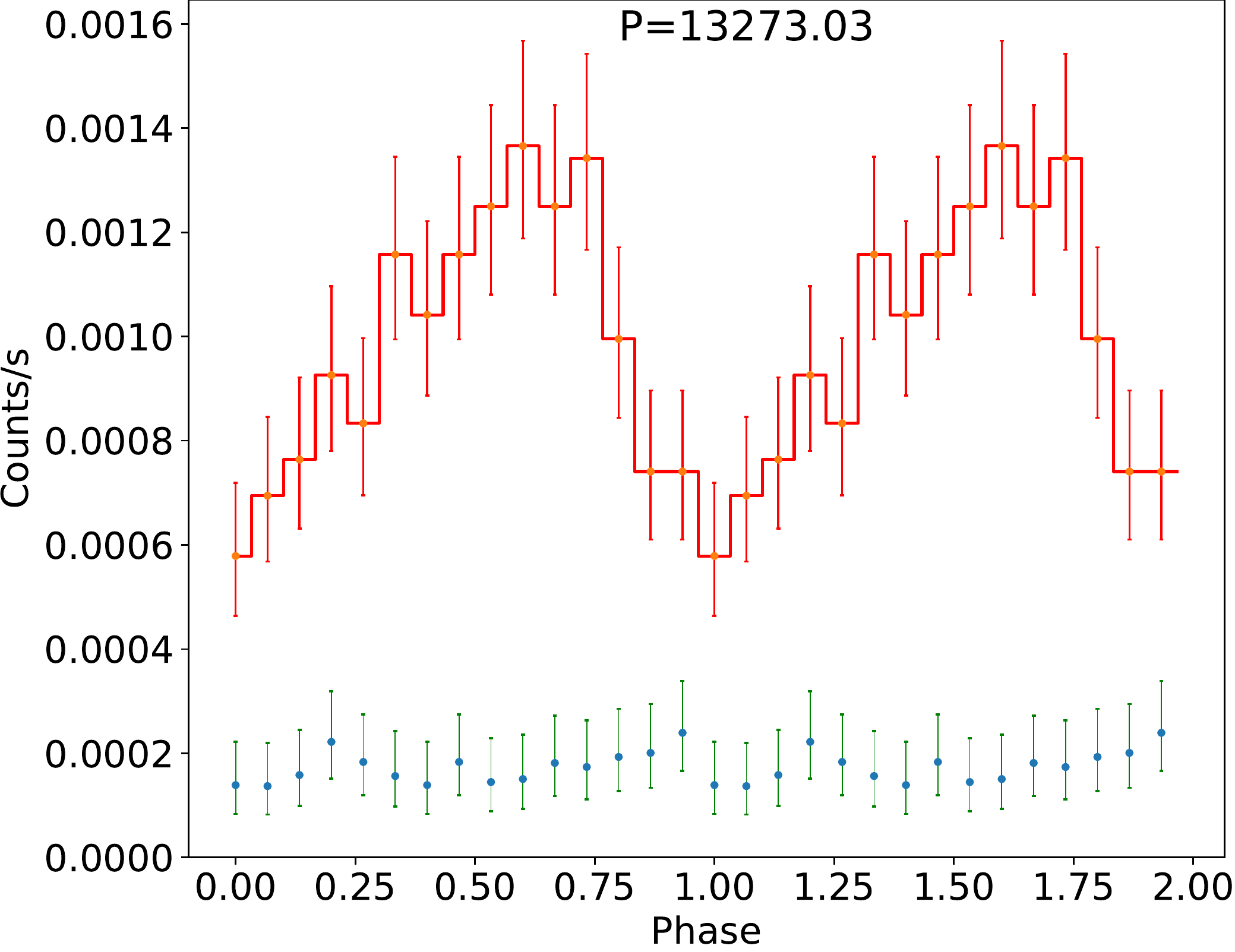}

\includegraphics[angle =0, width = 0.32\textwidth]{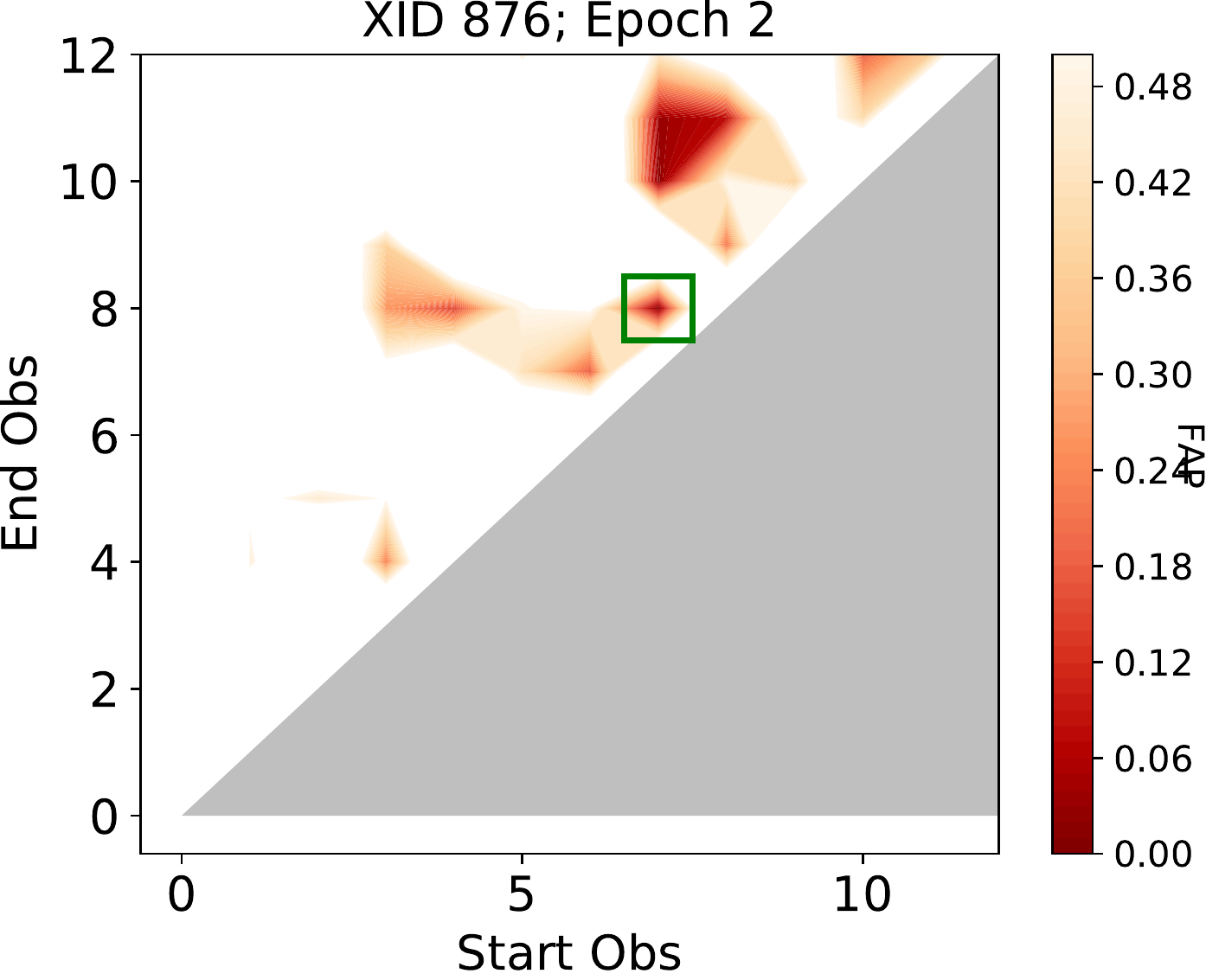}
\hfill
\includegraphics[angle =0, width = 0.32\textwidth]{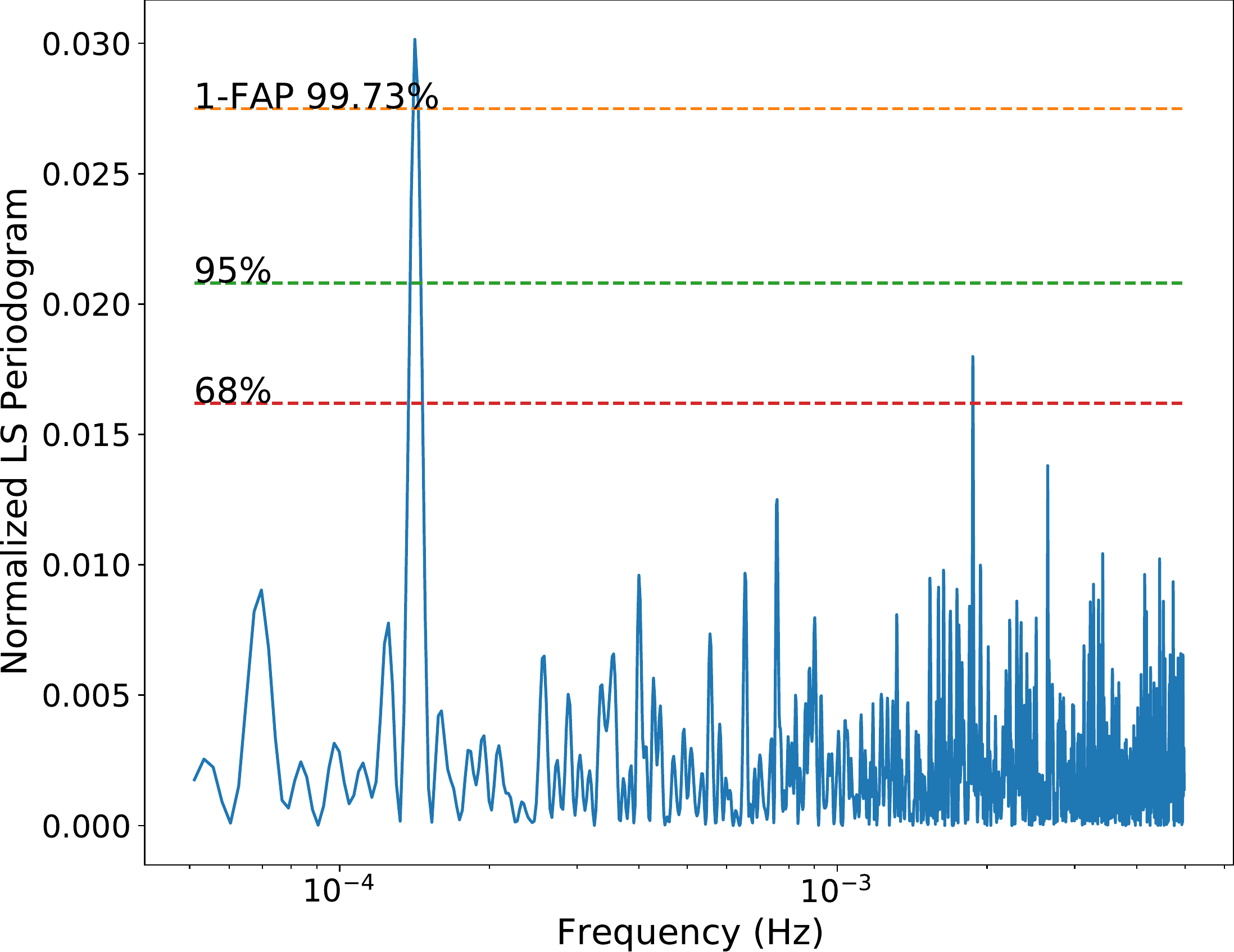}
\hfill
\includegraphics[angle =0, width = 0.32\textwidth]{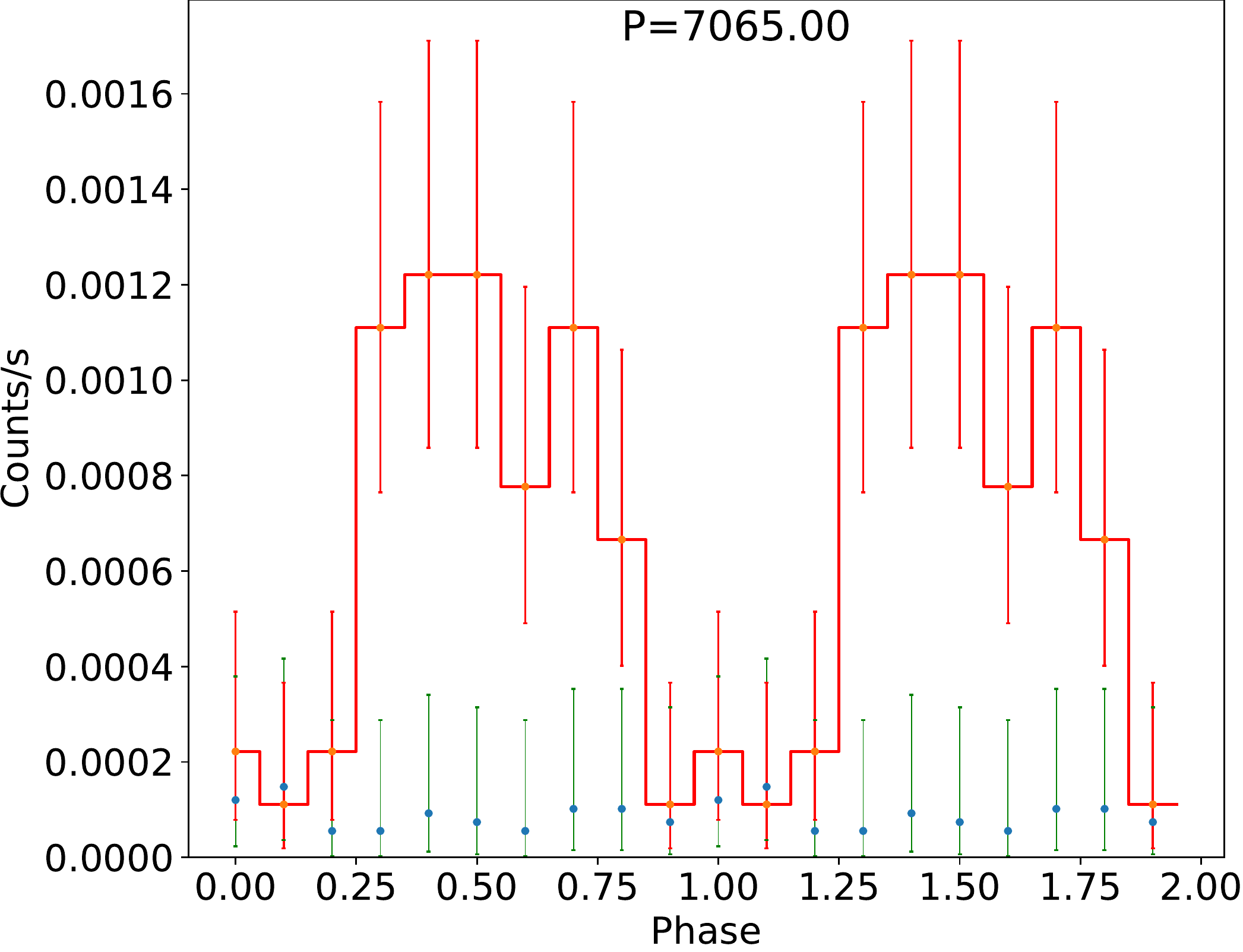}

\caption{Transient QPO candidates in source XID 643 ({\it upper row}) and XID 876 ({\it lower row}). 
{\it Left:} Distribution of FAP of the peak value of the LS periodogram, from light curves of all unique combinations of consecutive observations (defined by the start observation and end observation) within a given epoch. 
The white/green box marks the combination of consecutive observations with the lowest FAP value, i.e., the light curve possessing the most significant periodic signal. 
{\it Middle:} The LS periodogram from the light curve marked by the white/green box in left panel. The FAP of the signal is denoted. 
{\it Right:} Phase-folded light curve according to the peak frequency and based on the same combination of observations. The green error bars represent the local background level. 
}
\label{fig:candQPO}
\end{figure*}

Two sources exhibit a periodic signal with 1-FAP $>99.73\%$ from at least one light curve, as illustrated in Figure \ref{fig:candQPO}.
XID 643 (R.A.=53.143475, Decl.=-27.653576 [J2000]) has a mean count rate of $6.0 \times 10^{-4}{\rm~cts~s^{-1}}$ and exhibits a periodic signal at \textasciitilde 13273 s with a significance of 99.89\% among ten observations starting with ObsID 16620 and ending with ObsID 17573, lasting for about 3 months. 
Light curves from a subset of these ten observations also yield a a signal around the same period, although at somewhat lower significance.
As an additional check, the same light curve spanning the ten observations, when fed to the GL algorithm, gives a signal at a nearly identical period, but with a probability of only 74\%. 
This might be partly due to its location near the CCD edge in most observations of Epoch 4, which causes reduced sensitivity.
XID 643 is identified as a high-redshift  ($z$=1.51) AGN with an intrinsic luminosity of $9\times 10^{43}~{\rm erg~s^{-1}}$ from \citet{2017ApJS..228....2L}, suggesting an intrinsic quasi-period of $\sim$ 5288 s, if the signal was real.

XID 876 (R.A.=53.209310, Decl.=-27.881104 [J2000]) has a mean count rate of $5.5 \times 10^{-4}{\rm~cts~s^{-1}}$ and exhibits a statistically significant periodic signal of 7065 s (1-FAP = 99.92\%) in only one light curve, which is from one single observation (ObsID 8592) lasting 90.1 ks.  
The same light curve fed to the GL algorithm finds the same periodic signal with a probability of 99.84\%, leading some support to its reality. This source is identified as a high-redshift ($z$=3.47) AGN with an intrinsic luminosity of $4\times 10^{44}~{\rm erg~s^{-1}}$ from \citet{2017ApJS..228....2L}, suggesting an intrinsic quasi-period of $\sim$ 1581 s.

\section{Summary}
\label{sec:summary}
In this work we have presented the first systematic search for X-ray QPO among $\sim$ 1000 CDF-S AGNs. 
After dividing the 7-Ms {\it Chandra} observations into four epochs, we search for periodic signals that are persistent throughout any of these epochs, using the well-established methods of Lomb-Scargle periodogram and Gregory-Loredo Algorithm.
Among all the CDF-S sources, no statistically significant genuine periodic signal is found with either method on any of the four epochs.
Further dedicated simulations of source PSD and light curves suggest that this non-detection might be primarily due to a moderate sensitivity of the CDF-S data, in which only a small number of bright sources are present. 
With the help of simulation-predicted detection efficiency, we are able to provide a meaningful constraint on the intrinsic occurrence rate of AGN QPOs, $< (15-20)\%$, provided that they share a similar PSD shape with a handful of currently known cases and that they are persistent for months to years. 
The true intrinsic occurrence rate might be significantly below this upper limit, given the non-detection.
A robust determination of the QPO occurrence rate would require a much larger and homogeneous sample of bright AGNs. 

Our additional search for short-lived QPOs, referring to that only detected over a small subset of all observations, results in two statistically significant signals, one in source XID 643 at a period of $\sim$13273 s and the other in source XID 876 at a period of $\sim$7065 s.
The reality of these signals invite confirmation with future observations.

\section*{Acknowledgements}
This work is supported by the science research grants from the China Manned Spaced Project (CMS-CSST-2021-B02, CMS-CSST-2021-B11). 
The authors are grateful to Bin Luo for many valuable discussions, and to the anonymous referee for instructive suggestions that helped to improve our manuscript.

\section*{Data Availability}
The X-ray data and simulated light curves presented in this article will be shared on reasonable request to the corresponding authors.




\bibliography{example}{}
\bibliographystyle{mnras}




\appendix
\section{An analytic approximation of FAP}
\label{sec:FAP}

False alarm probability is a practical approach to assess the significance of a peak in the periodogram, based on the premise that the probability density function of the periodogram value is known.
For a normalized LS periodogram, the cumulative probability of observing a periodogram value less than $z$ at a single frequency should be
\begin{equation}
P_{\rm single}(z)=(1-z)^{\sqrt{\frac{N_{\rm bin}-3}{2}}},
\end{equation}
where $N_{\rm bin}$ is the number of time bins in the light curve.
Under the assumption that the value of the periodogram at each frequency is independent, 
the FAP can be estimated as,
\begin{equation}
{\rm FAP}(z) \approx 1-[P_{\rm single}(z)]^{N_{\rm eff}},
\end{equation}
where $N_{\rm eff}$ is the number of frequencies. A number of works have been devoted to provide an approximation of this value \citep{1986ApJ...302..757H,2004MNRAS.354.1165C,2008MNRAS.388.1693F,2008MNRAS.385.1279B}.
Among these, the \citet{2008MNRAS.385.1279B} method provides the most accurate approximation, 
\begin{equation}
\operatorname{FAP}(z) \approx 1-P_{\text {single}}(z) e^{-\tau(z)}	
\label{eqn:Baluev}
\end{equation}
where,
\begin{equation}
\tau(z) \approx \gamma_{\mathcal{H}} W\left(1- z\right)^{\frac{N_{\rm bin}-4}{2}} \sqrt{\frac{N_{\rm bin}}{2z}}.
\end{equation}
The order-of-unity factor $\gamma_{\mathcal{H}}$ can be neglected when $N_{\rm bin}$ is larger than 10. And $W=\nu_{\rm max} \sqrt{4\pi\mathbb{D} t}$, in which $\mathbb{D} t=\overline{t^{2}}-\overline{t}^2$ is the weighted variance of the time series and $\nu_{\rm max}$ is the maximum frequency under consideration.

%


\bsp	
\label{lastpage}
\end{document}